\documentclass[fleqn,twoside]{article}
\usepackage{espcrc2}

\usepackage{graphicx}
\usepackage[figuresright]{rotating}

\def\lsim{\raise0.3ex\hbox{$<$\kern-0.75em\raise-1.1ex\hbox{$\sim$}}}
\def\gsim{\raise0.3ex\hbox{$>$\kern-0.75em\raise-1.1ex\hbox{$\sim$}}}

\def\pom{{I\!\!P}}

\newcommand{\AmS}{{\protect\the\textfont2
  A\kern-.1667em\lower.5ex\hbox{M}\kern-.125emS}}

\title{Revisiting the phenomenology on the QCD color dipole picture}

\author{A.I. Lengyel\address[MCSD]{Institute of Electron Physics, National
Academy of Sciences of Ukraine, \\  Universitetska 21,
UA-88016 Uzhgorod, Ukraine } and  M.V.T. Machado \address{High Energy Physics Phenomenology Group, GFPAE IF-UFRGS, \\ Caixa Postal 15051, CEP 91501-970, Porto Alegre, RS, Brazil }}
       
\begin{document}

\begin{abstract}
 Using the QCD dipole picture of the hard BFKL Pomeron, we perform a 3 parameter fit analysis of the recent inclusive structure function experimental mesurements at small-$x$ and intermediate $Q^2$. As a byproduct, the longitudinal structure function and the gluon distribution are predicted without further adjustments. The data description is quite reasonable, being timely a further study using  resummed NLO BFKL kernels along the lines of recent theoretical developments.
\vspace{1pc}
\end{abstract}

\maketitle

\section{Introduction}

Deep inelastic electron-proton scattering (DIS) experiments at HERA have provided
measurements of the inclusive structure function $F_2(x,Q^2)$ in very
small values of the Bjorken variable $x \gsim 10^{-5}$. In
these processes the proton target is analyzed by a hard probe with virtuality
$Q^2=-q^2$, where $x \sim Q^2/2p.q$ and $p,\,q$ are the four-momenta of the
incoming proton and the virtual photon probe. In this domain, the
gluon is the leading parton driving the  small-$x$ behavior of the deep
inelastic observables. 

A sound  approach encoding all order  $\alpha_s\,\ln (1/x)$ resummation is the QCD dipole picture \cite{dipole}. It was proven that such approach reproduces the  BFKL evolution \cite{BFKL}. The main process is the onium-onium scattering, that is the reaction between two heavy quark-antiquark states (onia). This process is basically  perturbative due to the onium radius being the natural hard scale at which the strong coupling is computed. In the large $N_c$ limit, the heavy pair and the soft gluons are represented as a collection of color dipoles. The cross section is written as a convolution between the number of the dipoles in each onium state and the basic cross section for dipole-dipole scattering due to two-gluon exchange. The QCD dipole model can be applied to DIS process, assuming that the virtual photon at high $Q^2$ can be described by an onium. Furthermore, the proton is described by a collection of onia with an average onium radius to be determined from phenomenology. This model has produced a successful description of the old structure function data \cite{phenomenology}. These achievements are our main motivation in revisiting the QCD color dipole picture  and in applying it for description of the currently more accurate $F_2(x,Q^2)$ experimental results.

The approach also allows a systematic framework for testing the resummed next-to-leading order BFKL evolution kernels, producing predictions for the proton structure function. A method for doing this has been  proposed in Ref. \cite{Peschanski}, where the resummation schemes can  tested through the Mellin transformed $j$-moments of $F_2$. Moreover, it has been shown that a geometric scaling for the photon-proton cross section and the symmetry between low and high $Q^2$ regions are associated to the symmetry of the two-gluon dipole-dipole cross section \cite{munier1}. Furthermore, within the approach, a simple analytic expression for the dipole-proton scattering amplitude has been computed taking the scattering amplitude as a solution of the BFKL equation in the vicinity of the saturation line for dipole sizes $r$ (in the photon wavefuntion) obeying $r \lsim 1/Q_{\mathrm{sat}}$ \cite{IIM}, where $Q_s^2 (x) \propto e^{\,\lambda \,\log\,(1/x)}$ is the saturation scale. Finally, the approach has recently been  used to describe hard processes initiated by virtual-gluon probes \cite{dip_hadron}. 

In this contribution we summarize our fit analysis using the QCD dipole phenomenology applied to DIS process \cite{Lengyel_Machado_dipfit}. In what follows, the main expressions are presented and the fitting results are shown and discussed in conclusion. 

\section{The QCD dipole picture applied to DIS}

The starting point in the QCD color dipole picture is the onium-onium scattering. An onium is a  heavy quark-antiquark state, turning out the scattering  process  perturbative once the onium radius provides the hard scale entering into the strong coupling $\alpha_s$. As energy increases (small-$x$), these original onia can radiate soft gluons due to QCD evolution. In the large $N_c$ limit, the heavy pair and the soft gluons are represented as a collection of color dipoles. The cross section is then written as a convolution between the number of the dipoles in each onium state and the basic cross section for dipole-dipole scattering due to two-gluon exchange. The latter quantity is theoretically understood and perturbatively calculable. The physical process more suitable for using the approach is  $\gamma^*(Q_1) \,\gamma^*(Q_2)$ scattering, where the hard scale is provided by the virtualities of the virtual photons, $Q_{1,2}$. That is, the virtual photon at high $Q^2$ is assumed to be described by an onium with radius $r\sim 1/Q$. 

On the other hand, in the DIS process one has a two-scale problem where the hard scale is given by the photon virtuality and the soft one is associated to the proton typical size. Hence, the proton is approximately described by a collection of onia with an unknown average onium radius. Then, the DIS cross section is written as a convolution of the probability of finding an onium in the proton and the photon-onium cross section. Relying on renormalization group  properties, a  suitable ansatz for the former quantity was proposed \cite{phenomenology}. It depends on the average number of primary dipoles in the proton $n_{\mathrm{eff}}$ and on their average transverse diameter $r_0 \equiv 2/Q_0$. Under these assumptions and the convolution integral approximated by a steepest-descent method (using the expansion of the BFKL kernel near $\gamma=1/2$), the structure functions take a simple form \cite{phenomenology},
\begin{eqnarray}
F_{T,\,L} & = &  H_{T,\,L}\, \frac{\bar{\alpha}_s\,\pi^3 \,e_f^2\,n_{\mathrm{eff}}}{96}\,\left(\, \frac{x_0}{x} \, \right)^{\omega_{\pom}}\, \frac{Q}{Q_0}\nonumber \\
& \times &  \sqrt{2\,\kappa\,(x)/\pi}\,\exp \left[ -\frac{\kappa\,(x)}{2}\,\ln^2 \frac{Q}{Q_0} \right],
\label{sfs}
\end{eqnarray}
where $H_T=9/2$ and $H_L=1$. The hard Pomeron intercept is given by  $\alpha_{\pom}=1 + \omega_{\pom}$, with $\omega_{\pom}=4\,\bar{\alpha}_s \ln 2$ and  $\bar{\alpha}_s=\alpha_s N_c/\pi$. The BFKL diffusion coefficient at rapidity $Y=\ln \,(x_0/x)$ is written as $\kappa\,(x)=[\bar{\alpha}_s\, 7\, \zeta (3)\, \ln \frac{x_0}{x}]^{-1}$. The conditions to obtain Eq. (\ref{sfs}) from the saddle-point method constrain its region of applicability. Namely, the relation $\kappa \,(x)\,\ln (Q/Q_0)\ll 1$ should be obeyed. This is realized for the region of moderate $Q/Q_0$ when compared to the range on $x_0/x$. 

\section{Results and Conclusions}

Lets present the fitting procedure using the recent HERA experimental data  on the proton structure function \cite{H1rec,ZEUSrec} and taking  Eq. (\ref{sfs}), where  $F_2=F_T + F_L$. We  defined the overall normalization for $F_2$, ${\cal N}_p = (H_T + H_L)\,\bar{\alpha}_s\pi^3 e_f^2 \,n_{\mathrm{eff}}/96$. For the fit procedure we have considered only the small $x\leq 10^{-2}$ data, covering the range of virtualities $1.5 \leq Q^2 \leq 150$ GeV$^2$. We have also fixed $x_0=1$, since its value is  reasonably stable for different data sets. Therefore, we are left with a reduced number of parameters (${\cal N}_{p}$, $\alpha_{\pom}$, $Q_0$). The resulting parameters for H1 and ZEUS experimental data sets are presented in Table~\ref{table:1}. 

\begin{table}[t]
\caption{Parameters for H1 and ZEUS data sets \cite{H1rec,ZEUSrec}.}
\label{table:1}
\begin{center}
\begin{tabular}{||c|c|c||}
\hline
\hline
 $\mathrm{PARAMETER}$  &   ZEUS data set    &  H1 data set   \\
\hline
 ${\cal N}_p$ & 0.0977  &  0.0985  \\
  $Q_0$  & 0.571  &  0.587  \\
 $\alpha_{\pom}$ & 1.24  & 1.23 \\
\hline
\hline
$\chi^2/\mathrm{d.o.f.}$  & 1.08  &   1.02 \\
\hline
\hline
\end{tabular}
\end{center}
\end{table}

The quality of fit is quite good for both H1 and ZEUS data sets. We have performed also an extrapolation of the fit using in all range on $x$ (adding E665 the EMC data) and $Q^2$. In order to do this, a non-singlet contribution was added and large-$x$ threshold factors were considered. We quote Ref. \cite{Lengyel_Machado_dipfit}) for further details. The procedure presented here is similar to previous analysis on Refs. \cite{phenomenology}, with an even  lower effective power ($\alpha_{\pom}\simeq 1.282$ in \cite{phenomenology}). The low value for the fixed coupling constant $\alpha_s\simeq 0.1$ reveals the well known necessity of sizeable higher order corrections to the approach. Accurate analysis in this lines, considering resummed NLO BFKL kernels, has been proposed recently \cite{Peschanski} producing a reasonable $\alpha_s\sim 0.2$ for the typical $Q^2$ range considered in phenomenology for structure functions. 

Within the approach above, it is possible to determine longitudinal structure function $F_L$ without further adjustments. From Eq. (\ref{sfs}) and the definition of the overall normalization one has $F_L=(2/11)\,F_2$, since $H_T+H_L=11/2$ and $H_L=1$. The results are in good agreement with the recent data (see Fig. (3) in \cite{Lengyel_Machado_dipfit}). Moreover, in the QCD dipole approach the gluon distribution function can be calculated in a straightforward way. The result is  independent of the overall normalization, which contains part of the non-perturbative inputs of the model. The gluon distribution function is given by \cite{phenomenology,Lengyel_Machado_dipfit},
\begin{eqnarray}
x\,G(x,Q^2)  =   \frac{F_2(x,Q^2)}{\left[h_T\,(\gamma=\gamma_s)+ h_L\,(\gamma=\gamma_s) \right]}\,,
\label{xg}
\end{eqnarray} 
where $h_{T,L}(\gamma)$ are the LO photon impact factors, corresponding to the perturbative coupling to the photon \cite{phenomenology}. The quantity $\gamma_s = \frac{1}{2} \left[ 1 - \kappa\,(x)\,\ln \left( \frac{Q}{Q_0} \right)\right]$ comes from the  integration over $\gamma$ in the convolution of onia density inside the proton and photon-onium cross section. It is obtained via steepest descent method and taking the expansion of the LO BFKL kernel, $\chi (\gamma)$,  near $\gamma = \frac{1}{2}$, as referred before.

We have compared expressions Eq. (\ref{xg}) with NLO DGLAP resummed kernels. A sizeable deviation  between the recent H1 and ZEUS fits for that parton density function and the QCD dipole result is found \cite{Lengyel_Machado_dipfit}, whereas it is consistent with previous analysis. However, it should be stressed that the gluon distribution is an indirect observable, with  its determination being  model-dependent, and a direct comparison is not straightforward. We quote Ref. \cite{Lengyel_Machado_dipfit} for more details on this discussion.

As a final remark, very recently a pioneering phenomenological analysis using resummed NLO BFKL kernels in the saddle-point approximation has been done in Ref. \cite{PRS}. The LO fit provided paramaters similar to ours, with a slightly higher $\chi^2/\mathrm{d.o.f.}$.  The NLO fits give a qualitatively satisfactory account of the running $\alpha_s$ effect but quantitatively the quality of fit remains sizeably higher than the LO fit. This feature suggests the investigation of other  proposed theoretical resummation schemes and/or to improve those ones considered in the first analysis presented in Ref. \cite{PRS}.

\end{document}